\documentclass{aa} 

\bibpunct{(}{)}{;}{a}{}{,}
\usepackage[squaren,Gray]{SIunits}
\usepackage{color}
\usepackage{lmodern}
\usepackage{graphicx}
\usepackage{txfonts}
\usepackage{natbib,twoopt}
\usepackage[breaklinks=true]{hyperref}
\hypersetup{colorlinks=true,linkcolor=blue,citecolor=blue,urlcolor=cyan}
\makeatletter
\newcommandtwoopt{\citeads}[3][][]{\href{http://adsabs.harvard.edu/abs/#3}%
        {\def\hyper@linkstart##1##2{}%
                \let\hyper@linkend\@empty\citealp[#1][#2]{#3}}}
\newcommandtwoopt{\citepads}[3][][]{\href{http://adsabs.harvard.edu/abs/#3}%
        {\def\hyper@linkstart##1##2{}%
                \let\hyper@linkend\@empty\citep[#1][#2]{#3}}}
\newcommandtwoopt{\citetads}[3][][]{\href{http://adsabs.harvard.edu/abs/#3}%
        {\def\hyper@linkstart##1##2{}%
                \let\hyper@linkend\@empty\citet[#1][#2]{#3}}}
\newcommandtwoopt{\citeyearads}[3][][]%
{\href{http://adsabs.harvard.edu/abs/#3}
        {\def\hyper@linkstart##1##2{}%
                \let\hyper@linkend\@empty\citeyear[#1][#2]{#3}}}
\makeatother

\definecolor{a}{rgb}{0.90,0.40,0.10}

\begin{document}

\title{Cepheid distances from the SpectroPhoto-Interferometry of Pulsating Stars (SPIPS)}
\titlerunning{The SPIPS model for Cepheid distance determinations}
\subtitle{Application to the prototypes $\delta$~Cep and $\eta$~Aql}
\author{A. M\'erand\inst{1} \and P. Kervella\inst{2,3} \and J. Breitfelder\inst{1,2} \and A. Gallenne\inst{4} V. Coud\'e du Foresto\inst{2} \and T.~A.~ten~Brummelaar\inst{5} \and H.~A.~McAlister\inst{5} \and  S. Ridgway\inst{6} \and L.~Sturmann\inst{5} \and J.~Sturmann\inst{5} \and N.~H.~Turner\inst{5}}

 \institute{European Southern Observatory, Alonso de C\'{o}rdova 3107,
                        Casilla 19001, Santiago 19, Chile\\
                \email{amerand@eso.org}
                \and
                LESIA (UMR 8109), Observatoire de Paris, PSL, CNRS, UPMC, Univ. Paris-Diderot, 5 place Jules Janssen, 92195 Meudon, France
                \and
                        Unidad Mixta Internacional Franco-Chilena de Astronom\'{i}a, CNRS/INSU,
                        France (UMI 3386) and Departamento de Astronom\'{i}a, Universidad de Chile,
                        Camino El Observatorio 1515, Las Condes, Santiago, Chile
                        \and
                Universidad de Concepci\'{o}n, Departamento de Astronom\'{i}a, Casilla 160-C,
                        Concepci\'{o}n, Chile
                        \and
                        Center for High Angular Resolution Astronomy, Georgia State  
                    University, PO Box 3965, Atlanta, Georgia 30302-3965, USA
                    \and
                    National Optical Astronomical Observatory, Tucson, AZ, USA}
\date{Received ---; accepted ---}

  \abstract
   {The parallax of pulsation, and its implementations such as the Baade-Wesselink method and the infrared surface brightness technique, is an elegant method to determine distances of pulsating stars in a quasi-geometrical way. However, these classical implementations in general only use a subset of the available observational data.}
   {\citetads{2010ApJ...719..335F} suggested a more physical approach in the implementation of the parallax of pulsation in order to treat all available data. We present a global and model-based parallax-of-pulsation method that enables including any type of observational data in a consistent model fit, the SpectroPhoto-Interferometric modeling of Pulsating Stars (SPIPS).}
   {We implemented a simple model consisting of a pulsating sphere with a varying effective temperature and a combination of atmospheric model grids to globally fit radial velocities, spectroscopic data, and interferometric angular diameters. We also parametrized (and adjusted) the reddening and the contribution of the circumstellar envelopes in the near-infrared photometric and interferometric measurements. }
   {We show the successful application of the method to two stars: $\delta$~Cep and $\eta$~Aql. The agreement of all data fitted by a single model confirms the validity of the method. Derived parameters are compatible with publish values, but with a higher level of confidence.}
   {The SPIPS algorithm combines all the available observables (radial velocimetry, interferometry, and photometry) to estimate the physical parameters of the star (ratio distance/$p$-factor, T$_\mathrm{eff}$, presence of infrared excess, color excess,
etc). The statistical precision is improved (compared to other methods) thanks to the large number of data taken into account, the accuracy is improved by using consistent physical modeling and the reliability of the derived parameters is strengthened thanks to the redundancy in the data.}


   \maketitle
%

\section{Introduction}

Cepheids are the backbone of the extragalactic distance ladder because their pulsation periods, which are easily measured observationally, correlate directly with their luminosities through Leavitt's law (the period-luminosity relation, \citeads{1908AnHar..60...87L}; \citeads{1912HarCi.173....1L}). Thanks to their very high intrinsic brightness, they are visible in distant galaxies, as demonstrated for instance by \citetads{2001ApJ...553...47F} or \citetads{2011ApJ...730..119R}. They overlap with secondary, far-reaching distance indicators, such as type Ia supernovae (SN Ia) or the Tully-Fischer relation, whose scales are anchored to Cepheid luminosities.
Direct distance estimation of nearby Cepheids plays a crucial role in the calibration of Leavitt's law and, as a consequence, of the extragalactic distance ladder used to observationally estimate the Hubble constant $H_0$ \citepads[e.g.][]{2011ApJ...730..119R}.
This importance has recently been reaffirmed by \citetads{2012arXiv1202.4459S}: to the question \textit{``Are there compelling scientific reasons to obtain more precise and more accurate measurements of $H_0$ than currently available?''}, the authors answered \textit{``A measurement of the local value of $H_0$ to one percent precision (i.e. random errors) and accuracy (i.e. systematic errors) would provide key new insights into fundamental physics questions and lead to potentially revolutionary discoveries.''}  These authors also recognized the role of the Cepheids and the problem of controlling the systematics in their distance determinations.
An elegant and powerful method of directly measuring distances to Cepheids is the parallax of pulsation, also known as the Baade-Wesselink (BW) method \citepads{1926AN....228..359B, 1946BAN....10...91W}, although \citetads{1918MNRAS..78..639L} suggested the same method eight years earlier, but has never been credited for it.
In the BW technique, the variation of the angular diameter $\theta$ is compared to the variation of the linear radius (from the integration of the pulsation velocity $V_\mathrm{puls}$). The distance $d$ of the Cepheids is then obtained as the ratio between the linear and angular amplitudes,
\begin{center}
\begin{equation}
\theta(t)-\theta(0) \propto \frac{1}{d} \int_{0}^{t}V_\mathrm{puls}(\tau)d\tau
\label{eq:pop}
.\end{equation}
\end{center}

The BW method uses in practice a combination of two quantities: (1) disk-integrated radial velocities, estimated from the changing Doppler shift of photospheric absorption lines, and (2) angular diameters, either derived from multicolor photometric measurements and surface brightness relations, or from interferometric measurements.
One common property of these quantities is that they are derived from observations using models or some physical assumptions, therefore breaking the geometric nature of the parallax of pulsation.
The BW method has demonstrated its capability to reach the one-percent statistical precision regime \citepads[e.g.,][]{2005AandA...438L...9M}, and its true current limitation lies in the systematic uncertainties,
which are probably between five and ten percent. Two problems directly contribute to these systematics: the projection factor $p$ and the presence of circumstellar envelopes (CSEs).
The projection factor is a multiplicative correction factor applied to the radial velocity derived from a spectroscopic absorption-line Doppler shift. This factor is used to unbias the spectroscopic measurement and estimate the true pulsation velocity. To first order, the radial velocity can be seen as the projection of the pulsation velocity, integrated over the surface of the star. Since the pulsation of Cepheids is radial, the limb of the star does not have a Doppler shift, whereas the point at the center of the apparent stellar disk has a maximum projected velocity toward the observer.
Assuming a pulsation velocity of 1\,km/s, the measured disk-integrated radial velocity would be $1/p=1/1.5=0.67$\,km/s for a uniformly bright sphere. $p$ is lower than 1.5 for a limb-darkened star and more than 1.5 for a limb-brightened star. The p-factor is important because it biases the derived distance linearly: $d/p$ is the unbiased measurement in the parallax of pulsation equation (Eq.\ref{eq:pop}). 
For a long time, the adopted values of $p$ were based on the linear period-$p$-factor relation established by \citetads{1986PASP...98..881H, 1989ApJ...341.1004H}: $p=1.39 - 0.03 \log P$. This gives a value of $p \approx 1.36$ for a typical ten-day-period Cepheid, which was the most commonly used value in the literature (see, e.g., \citeads{1986AandA...168..139B}). But with the first direct determination of the p-factor of $1.27\pm0.06$ for the star $\delta$~Cep \citepads{2005AandA...438L...9M}, there has been a renewed interest in estimating the value of $p$. This work was based on the availability of a geometrical distance measurement, using the Fine Guidance Sensor (FGS) of the Hubble Space Telescope (HST). Since then, a dozen Cepheids have had their parallax measured directly in the same fashion \citepads{2007AJ....133.1810B}. This allows us to estimate more values of $p$, and even calibrate it as a function of the pulsation period, using the infrared surface brightness (IRSB) version of the parallax-of-pulsation method \citepads{2011AandA...534A..94S}. Stars are limb-darkened in the spectral continuum and more darkened at shorter wavelength. However, it should be noted that stellar surfaces are slightly limb-brightened inside absorption lines. This leads to an apparent paradox: one would expect the $p$-factor to be 1.5 or higher, even though direct measurements instead
lead to values of around 1.3.

To avoid the need of calibrating the projection factor, another approach is to include its contribution in the pulsation model. In their recent work, \citetads{2007PASP..119..398G} attempted to directly extract the pulsation velocity by using a simple geometric model of an absorption line deformed by the pulsation: the resulting $p$-factors they found for the radial velocity published using different measurement techniques vary from 1.30 to 1.38 for given star, leading to a systematic error of 6\% on the parallax of pulsation distances. Again, this value is for a given star and results from the various data-reduction techniques (e.g., bisector, cross-correlation) used to extract the radial velocity from spectra \citepads{2009AandA...502..951N}.
Another potential source of bias is the presence of circumstellar envelopes, which have been discovered and studied in the infrared by \citetads{2006AandA...448..623K}, \citetads{2006AandA...453..155M, 2007ApJ...664.1093M}, \citetads{2009AandA...498..425K}, and \citetads{2011AandA...527A..51G,2012AandA...538A..24G,2013AandA...558A.140G}. In the context of the parallax of pulsation, these envelopes affect the infrared apparent brightness of the star from the K-band ($2\,\mu$m) and longward of this. They also bias the angular diameters measured by infrared long-baseline interferometry. The geometry of the CSE seems to be almost universal (\citeads{2006AandA...448..623K}; \citeads{2006AandA...453..155M,2007ApJ...664.1093M}) and to vary only in intensity.
Even in the \textit{Gaia} era, when a few hundred Galactic Cepheids will have their distance measured accurately, the parallax of pulsation will still be a invaluable tool for distance investigation. One might think, for instance, of studying the Large Magellanic Cloud Cepheids using this technique. In addition, it should be noted that the parallax of pulsation will remain an important tool for studying the physics of Cepheids: \textit{Gaia} providing the distances, the BW studies of Galactic Cepheids will investigate the physics which it relies on.
\section{Integrated method}

\subsection{Motivations}

This work is the natural evolution of the method suggested by \citetads{1976MNRAS.174..489B} to estimate the angular diameter from photometry. The generalization of the idea was proposed by  \citetads{2010ApJ...719..335F} to provide a better physical basis for the parallax of pulsation and to call for taking into account all possible observables. They proposed to use a universal surface brightness to compute magnitudes, based on the following formula (for example, for band B):
\begin{equation}
B = B_0 - C_B\times\log{T_\mathrm{eff}} - 5\log{\theta} + A_B\times E(B-V)
\label{eq:phot}
,\end{equation}
where $\theta$ is the Rosseland angular diameter, $T_\mathrm{eff}$ the effective temperature, $E(B-V)$ the color excess, $B_0$ and $C_B$ a set of parameters describing the surface brightness relation, and $A_B$ the bandpass-dependent reddening coefficient. This method has the disadvantage of requiring a calibration of $B_0$ and $C_B$, and, more important, assumes a dependency of the surface brightness (here, a linear relation in effective temperature). These relations were recently calibrated by \citetads{2012ApJ...748..107P} by analyzing thousands of measurements for dozens of Cepheids. 
We propose to use a different method that is unique thanks to a combination of two things:
\begin{itemize}
\item We propose a "fit all at once" method (for a given star), which takes into account all the observables and fit all the parameters. This has the advantage of offering the best statistical accuracy and confidence in the result. Usually, BW methods are implemented by steps: first a radial velocity function is fitted analytically, then it is integrated, and finally compared to the angular diameter measurements to derive the distance. Unless treated properly (using a bootstrapping method, for example), this leads to an underestimation of the uncertainty of the final distance, unless the uncertainties on prior steps of the methods are propagated properly (e.g., the uncertainty on the radial velocity Fourier fit).\item We try, as much as possible, to physically model the observables. For example, we propose synthesizing photometry based on atmospheric models and using calibrated bandpass filters, instead of using analytical surface brightness relations linear in color (such as V-K), which we know are not observationally linear, see for example \citetads{2004A&A...428..587K}. 
\end{itemize}
This approach also offers the potential of investigating, for example, why, in the case of $\delta$~Cep, the interferometric angular diameters of \citetads{2005AandA...438L...9M} and the angular diameters derived by IRSB by \citetads{2012AandA...543A..55N} seem to systematically disagree by about 4\%. A global method should be able to provide an answer to this contradiction. Another advantage of such a method is also to relax the constraint of uniform phase coverage to a certain extent; this was previously recognized by \citetads{2010ApJ...719..335F}. 

It is remarkable that global methods using physics-based models are quite widespread in the field of determining fundamental parameters of eclipsing binaries. Implementations such as PHOEBE\footnote{\url{http://phoebe-project.org/}} \citepads{2005ApJ...628..426P} or ROCHE \citepads{2012IAUS..282..279P} use the same philosophy as we mentioned above. As a first path to implement such a method for Cepheids (this work), we developed a global approach for deriving fundamental parameters of the eclipsing binary $\delta$~Vel \citepads{2011A&A...532A..50M}, which we successfully checked against the ROCHE model of the same system \citepads{2011A&A...528A..21P}.

\subsection{Description of the model}
We assumed that Cepheids are radially pulsating spheres, with perfect cycle-to-cycle repetition of their physical properties. The pulsation velocity and the effective temperature as a function of phase are described by periodic functions of the pulsation phase $\phi$, interpolated using splines or Fourier series. \citetads{2005AandA...438L...9M} showed that periodic spline functions often offer a better description of the pulsation of Cepheids than do Fourier series, since Cepheids often exhibit pulsation velocity variations that are very different from a simple sinusoidal wave. This requires many Fourier harmonics to describe the pulsation profile properly. Additionally, Fourier series fits are very sensitive to poor phase coverage and tend to introduce non-physical oscillations. This means that Fourier decomposition requires a very uniform and dense phase coverage, which is not always available. However, Fourier series offer a good numerical stability, which is not always the case for a spline with free-floating nodes. In practice, we implemented both methods to allow for more flexibility. By default, Fourier series are used because they allow quicker computation and certain numerical convergence. We then switched to splines and kept this option if the goodness of fit was improved. Another important assumption was that Cepheid photospheres can be approximated by hydrostatic models in terms of energy distribution and center-to-limb darkening. We used the set of astrophysical constants recently recommended by \citetads{2011PASP..123..976H}.
\paragraph{Atmospheric models:} To compute synthetic photometry, we used ATLAS9 atmospheric models\footnote{\url{http://wwwuser.oats.inaf.it/castelli/grids.html}}, with solar metallicity and a standard turbulent velocity of 2km/s. The effect of metallicity on the magnitudes is very weak, as noted by \citetads{2014MNRAS.444..392C}. We used a grid of models spaced by 250K in effective temperatures and by 0.5 in logg. In practice, for each photometric bandpass, we reduced the models to a grid of magnitudes computed for an angular diameter of 1\,mas. We then modeled the photometry by using the formula (here in B band)
\begin{equation}
B = B_\mathrm{\theta=1mas}(T_\mathrm{eff}, \mathrm{logg}) - 5\log{\theta} + A_B\times E(B-V).
\end{equation}
This equation is similar to Eq.~\ref{eq:phot}, except that the linear surface brightness relation is replaced by a grid of interpolated values $B_\mathrm{\theta=1mas}$, which is a function of the model: $T_\mathrm{eff}$ and $\mathrm{logg}$. $T_\mathrm{eff}(\phi)$ is fitted to the data (using either splines or Fourier series). On the other hand, $\mathrm{logg}$ is deduced from the parameters of the model: the mass of the star is assumed using the period-radius-mass relation of \citeads{2001ApJ...563..319B}, and the linear radius is known internally in the model. The sensitivity of the $M_\mathrm{\theta=1mas}$ to the gravity is, in any case, very low: this means that the choice of mass for the model is quite unimportant. As noted by \citetads{2014MNRAS.444..392C}, atmospheric models are poorly suited for reproducing synthetic photometry bluer than the B band, hence we limit our modeling to a range of 0.4$\mu$m (B band) to about 2.5$\mu$m (K band): the data presented here used the Johnson system in the visible (B and V bands), as well as the Walraven system (B and V band) and the CTIO system in the near-infrared (J, H, and K bands).
\paragraph{Photometric bandpasses and zero-points:} The photometric magnitudes were computed for each model of the grid, using band-passes and zero-points from the Spanish Virtual Observatory (SVO) database\footnote{\url{http://svo2.cab.inta-csic.es/theory/fps3/} and \url{http://www.ivoa.net/documents/Notes/SVOFPS/}} and the Asiago Database on Photometric System\footnote{\url{http://ulisse.pd.astro.it/Astro/ADPS/Paper/index.html}} \citepads{2000A&AS..147..361M} for the Walraven systems. Note that in the case of Walraven, we multiplied all the magnitudes by -2.5 since this unusual system expresses magnitude as the logarithm of the flux, without using the conventional -2.5 multiplicative factor. This allows for a uniform numerical treatment of all the photometric measurements. For the zero points, we chose the filters in the SVO that were recently calibrated by 
\citeads{2014arXiv1412.1474M} (see Table~\ref{Tab:filters}).
\begin{table*}
  \caption{Adopted filters and zero points}
  \label{Tab:filters}
\begin{center}
  \begin{tabular}{lrcccc}
  \hline
  filter &  $\lambda_\mathrm{eff}$ & zero point & SVO FilterID & Note & ref\\
  & (nm) & (W.m$^{-2}$.$\mu\mathrm{m}^{-1}$) \\ 
  \hline
B$_\mathrm{T}$ & 422.0 & $6.588\times10^{-08}$ & TYCHO/TYCHO.B\_MvB & revised by MvB 2014 &(1) \\
B$_\mathrm{W}$ & 432.5 & $1.230\times10^{-10}$ & --- & -2.5 Walraven filter B &(2) \\
B & 436.5 & $6.291\times10^{-08}$ & GCPD/Johnson.B & revised by MvB 2014 &(1) \\
B$_\mathrm{ST}$ & 466.7 & $5.778\times10^{-08}$ & GCPD/Stromgren.b & revised by MvB 2014 & (1) \\
HP & 517.1 & $3.816\times10^{-08}$ & Hipparcos/Hipparcos.Hp\_MvB & revised by MvB 2014 &(1) \\
V$_\mathrm{T}$ & 525.8 & $3.946\times10^{-08}$ & TYCHO/TYCHO.V\_MvB & revised by MvB 2014 &(1) \\
V$_\mathrm{W}$ & 546.7 & $6.730\times10^{-11}$ & --- & -2.5 Walraven filter V &(2) \\
V & 545.2 & $3.601\times10^{-08}$ & GCPD/Johnson.V & revised by MvB 2014 &(1) \\
Y$_\mathrm{ST}$ & 546.5 & $3.625\times10^{-08}$ & GCPD/Stromgren.y & revised by MvB 2014 & (1) \\
R & 643.7 & $2.143\times10^{-08}$ & GCPD/Cousins.R & revised by MvB 2014 &(1) \\
J & 1240.0 & $3.052\times10^{-09}$ & CTIO/ANDICAM.J &  &(1) \\
H & 1615.3 & $1.200\times10^{-09}$ & CTIO/ANDICAM.H &  &(1) \\
K & 2129.9 & $4.479\times10^{-10}$ & CTIO/ANDICAM.K &  &(1) \\
 \hline
  \end{tabular}
\tablefoot{ $^{(1)}$ Spanish Virtual Observatory; $^{(2)}$ "The Asiago Database on Photometric Systems"\citepads{2000A&AS..147..361M}; MvB 2014 refers to \citeads{2014arXiv1412.1474M}. }
\end{center}
\end{table*}
\paragraph{Reddening:} We parametrized the interstellar reddening using the B-V color excess, E(B-V), and the reddening law from \citetads{1999PASP..111...63F}, taken for Rv=3.1. Because the correction depends on the spectrum of the observed object, we computed all our reddening corrections using a template spectra for actual effective temperature at the phase at which the photometric observations were made. Reddening values for $T_\mathrm{eff}$=4500K, 5500K, and 6500K are listed in Table~\ref{Tab:reddening} for the various photometric systems we used. This is significantly different from traditional BW implementation. Reddening correction factors $R_\lambda$ are usually computed for Vega, a star much hotter than the Cepheids. For example, \citetads{2007AandA...476...73F} quotes $R_V$ (i.e., for the V band) values between 3.10 and 3.30 and adopted a value of 3.23. As seen in our Table~\ref{Tab:reddening}, our value for V$_\mathrm{GCPD}$ (Johnson) ranges from 3.00 to 3.05 between T$_\mathrm{eff}$=4500K to T$_\mathrm{eff}$=6500K (it would be 3.1 for T$_\mathrm{eff}$=10,000K). We note that the effect of our choice of computation of the reddening is most notably different for blue filters and makes the least difference for the near-infrared K-band.
Our choice of Rv=3.1 is mostly based on consensus and does not play a important role in the result: as far as we are concerned, the degeneracy is one-to-one between the reddening law Rv and the color excess E(B-V). In other words, changing the fixed value of Rv changes the fitted value of E(B-V) and maintains the other parameters of the fit within their fitted values.
\begin{table}
  \caption{Subsets of magnitudes for $\theta=1mas$ and reddening law (for Rv=3.1) for 3 values $T_\mathrm{eff}$ and logg=1.5}
  \label{Tab:reddening}
\begin{center}
  \begin{tabular}{lcc}
  \hline
  filter  &  $M_\mathrm{\theta=1mas}$, logg=1.5 & A$_\lambda$\\
  &  T$_\mathrm{eff}$=4500, 5500, 6500K & T$_\mathrm{eff}$=4500, 5500, 6500K \\
  \hline
B$_\mathrm{T}$ & 7.734, 5.799, 4.428 & 4.086, 4.146, 4.179 \\
B$_\mathrm{W}$ & 0.759, -1.132, -2.460 & 4.071, 4.101, 4.117 \\
B & 7.372, 5.625, 4.363 & 3.869, 3.954, 4.012 \\
B$_\mathrm{ST}$ & 6.890, 5.321, 4.219 & 3.800, 3.803, 3.805 \\
HP & 6.276, 5.018, 4.114 & 2.836, 2.990, 3.130 \\
V$_\mathrm{T}$ & 6.261, 4.950, 4.034 & 3.127, 3.173, 3.207 \\
V$_\mathrm{W}$ & -0.696, -1.967, -2.852 & 3.041, 3.058, 3.071 \\
V & 6.126, 4.864, 3.986 & 2.996, 3.027, 3.050 \\
Y$_\mathrm{ST}$ & 6.118, 4.855, 3.974 & 3.048, 3.053, 3.056 \\
R & 5.515, 4.467, 3.768 & 2.346, 2.371, 2.393 \\
J & 4.159, 3.602, 3.244 & 0.802, 0.804, 0.805 \\
H & 3.605, 3.280, 3.082 & 0.525, 0.527, 0.528 \\
K & 3.472, 3.217, 3.043 & 0.354, 0.354, 0.354 \\
\hline
  \end{tabular}
\end{center}
\end{table}
\paragraph{Center-to-limb darkening:} The effect of the center-to-limb darkening (CLD) needs to be taken into account to properly interpret interferometric angular diameters. Interferometers do not measure diameters directly, they measure visibilities, which need to be modeled in order to estimate an angular diameter. This is easiest to do using a uniform disk (UD) model. However, the derived diameter is not the true stellar diameter. Many authors have published tables of diameter corrections UD/LD, but we found that none are satisfactory, for the simple reason that the UD/LD correction depends on the spatial frequency at which the observations were made, because of the slight difference between UD and LD visibility profiles. For this reason we computed our own $\theta_\mathrm{UD}/\theta_\mathrm{Ross.}$ corrections.

The truly interesting radius in our case is the bolometric radius, which almost matches the  Rosseland value (where the average optical depth is 1). The Rosseland radius is the one that enters in the identity $L_\mathrm{bol}\propto R_\mathrm{Ross.}^2T_\mathrm{eff}^4$ \citepads{1991A&A...246..374B}. In the context of this work, we used a grid of photospheric models tabulated in effective temperature: this is why the apparent Rosseland diameter ($\theta_\mathrm{Ross}$) is the one that allows to compute accurate synthetic photometry.

\begin{figure}
  \centering
  \includegraphics[width=0.48\textwidth]{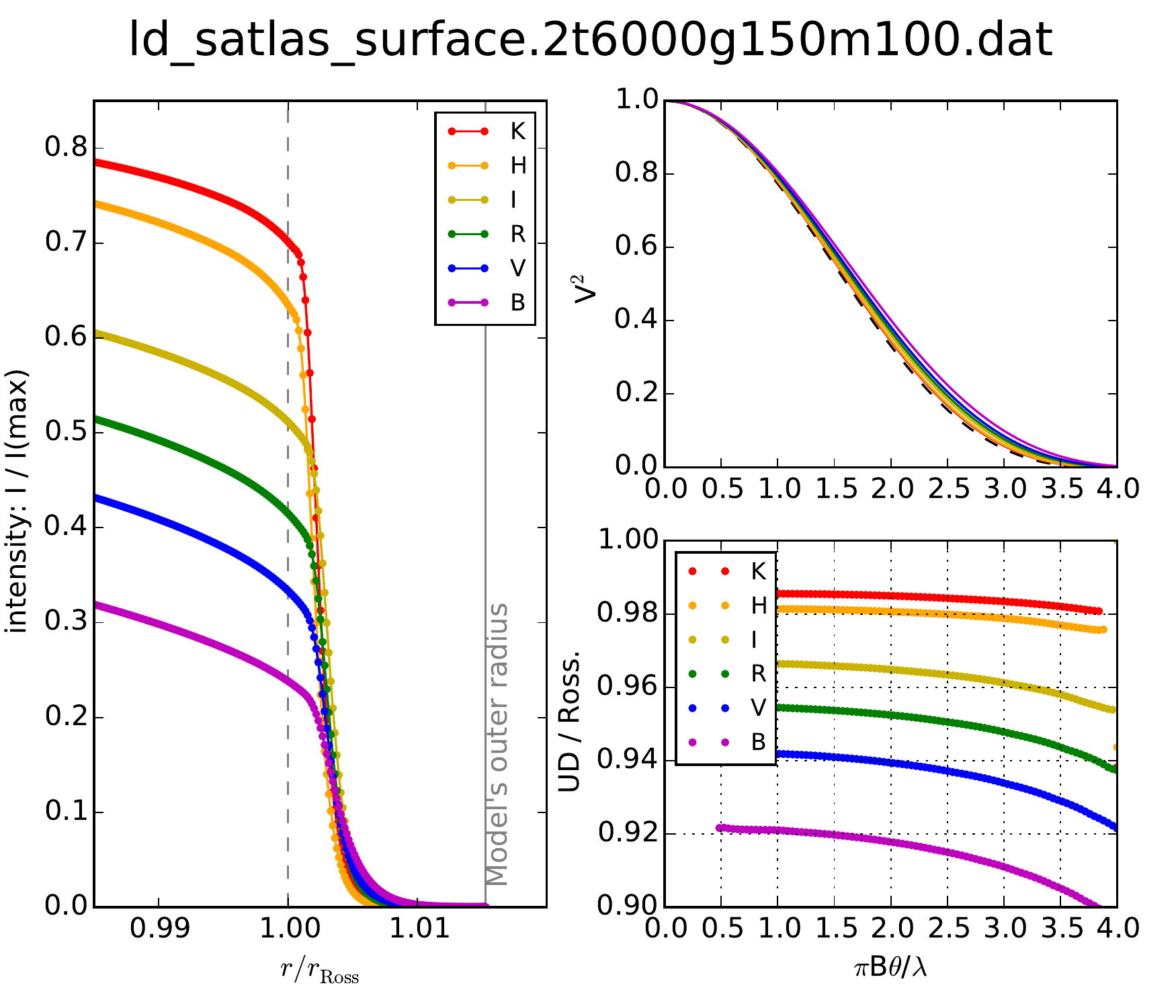}
  \caption{Example of deriving the interferometric correction factor $\theta_\mathrm{UD}/\theta_\mathrm{Ross.}$ for SATLAS model T$_\mathrm{eff}$=6000K, logg=1.5 and M=10\;M$_\odot$. \textbf{Left:} radial intensity profile, close to the limb ($\pm1\%$), for various bands; \textbf{upper right:} corresponding visibility functions as a function of the dimensionless spatial frequency $x = \pi B \theta/\lambda$; \textbf{lower right:} corresponding factors $\theta_\mathrm{UD}/\theta_\mathrm{Ross.}$ for each band as a function of $x$. }
  \label{fig:Ross}%
\end{figure}

We did not use ATLAS models for our own CLD correction because these models are plane-parallel and cannot produce accurate CLD profiles. Instead, we used grids of SATLAS models in the Cepheid range (\citeads{2013A&A...554A..98N}). The actual CLD profiles are available in the Vizier database (\citeads{2013yCat..35549098N}, via FTP\footnote{\url{ftp://cdsarc.u-strasbg.fr/pub/cats/J/A\%2BA/554/A98/spheric/}}). We extracted the radial intensity profile I(r), which was converted to a visibility profile using a Haenkel transform, for various spatial frequencies (expressed as $x = \pi B \theta/\lambda$, where B is the baseline in meters, $\theta$ the angular diameter in radian and $\lambda$ the wavelength in meters). For each spatial frequency, we scaled the spatial frequency of a uniform disk visibility profile to match the synthetic profile: the scaling factor was the ratio $\theta_\mathrm{UD}/\theta_\mathrm{Ross}$. An example is shown in Fig.~\ref{fig:Ross}. We note that spherical models, tabulated as I($\mu$) (where $\mu=\sqrt{1-r^2}$), do not have their limb for $r=1$, in contrast to plane-parallel models. This is because for spherical models, $r=1$ is the outer boundary of the model (defined as the optical depth in the case of SATLAS, \citeads{2008A&A...490..807N}) and does not correspond to the Rosseland radius. We used a separate tabulation of $R_\mathrm{Rosseland}/R_\mathrm{outer}$ extracted from the grid of SATLAS models (H. Neilson, private communication). The mathematical justification of the equivalence of the scaling in \textit{r} in the intensity profile and scaling the visibility curve to estimate the unbiased Rosseland angular diameter is a fundamental property of the Fourier transform: $V[I(a\times r), B\theta_{LD}] = V[I(r), B\theta_{LD}/a] = V[I(r), B\theta_{Ross.}], $ where $B$ is the baseline and $a=1/r_\mathrm{Ross.}$

We note that our results notably depart from those of \citetads{2012A&A...541A.134N} for two reasons: 1) we took the radius of the star as the Rosseland radius, not the outer layer of the SATLAS model (defined as $\theta_\mathrm{LD}$ by \citeads{2012A&A...541A.134N}), and 2) our $\theta_\mathrm{UD}/\theta_\mathrm{Ross}$ is a function of angular diameter and baseline. Overall, we found our values of $\theta_\mathrm{UD}/\theta_\mathrm{Ross}$ to be higher than those published in \citetads{2012A&A...541A.134N}. 

A limitation of our approach is that we used hydrostatic atmospheric models to compute our UD/Rosseland correction. This is not the latest way, since \citeads{2003ApJ...589..968M} have used updated models to take into account non-hydrostatic effects. These authors found that the UD/Rosseland correction is, on average, comparable with the hydrostatic values and that the variation of the correction, due to the pulsation, is very small: about 0.3\% in the near-infrared and up to 1.5\% in the visible. This translates more or less into the same bias in $d/p$ bias. Since we mostly
used near-infrared optical interferometric data, the bias from our choice of using hydrostatic models is, to the best of our knowledge, only about 0.3\%, at most. Moreover, there are no published grids of hydrodynamic models.

\begin{figure}
  \centering
  \includegraphics[width=0.48\textwidth]{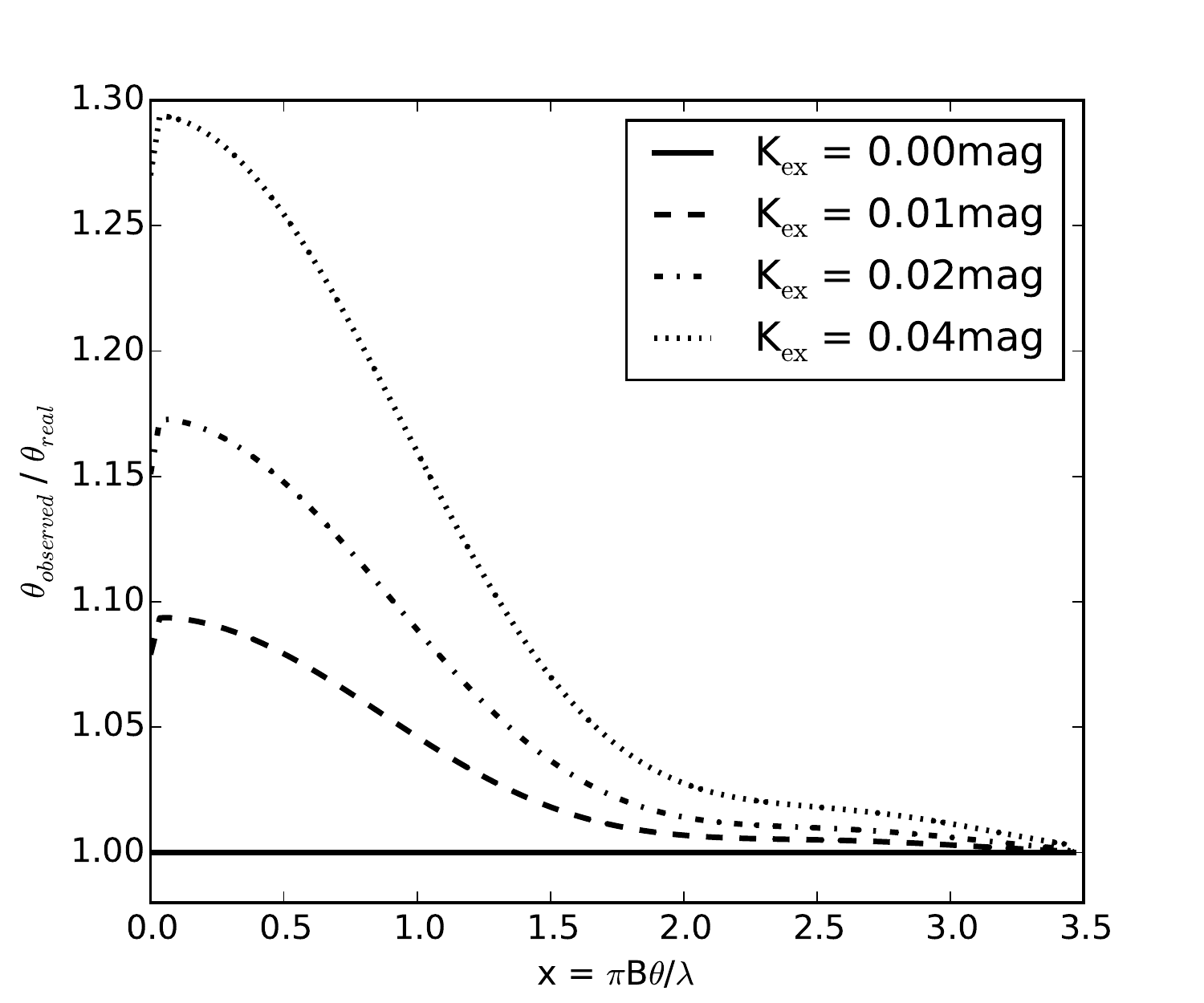}
  \caption{K-band interferometric angular diameter bias (observed / real) due to the CSE as a function of the dimensionless spatial frequency.}
  \label{Fig:bias}%
\end{figure}

\paragraph{Circumstellar envelopes:} The CSEs have two observational effects. The first one is on the near infrared photometric measurements, which are potentially biased for wavelengths in the K band ($2.2\,\mu$m) and redder. The second effect is on the interferometric angular diameters. \citetads{2006AandA...448..623K} and \citetads{2006AandA...453..155M} showed that the fringe visibility as a function of the baseline length departs from the classical function of a limb-darkened star. In the case of the CSE, the bias on the measurements depends on the baselines and angular diameter. The approach we adopted was to use a grid of models using the parametrization reported
by \citetads{2005A&A...436..317P}, allowing the tabulation of the angular diameter bias as a function of infrared excess. Biases ($\theta_{observed}$ / $\theta_{real}$) for different strengths of CSEs are shown on Fig.~\ref{Fig:bias}. We also allowed for an excess in H band, since these two bands are relatively close in wavelengths and it is hard to imagine that the CSEs produce K-band excess and no H-band excess. If no H excess is given as an input parameter, we chose to consider an H-band excess twice as low as the K band excess. The numerical process is very similar to the one we described for the limb-darkening correction: we synthesized the visibilities of a limb-darkened disk surrounded by the CSE, with the relevant observational parameters, and we fitted a uniform disk model to estimate the bias. This is numerically costly, but it is the only accurate way to estimate the bias.

\subsection{Fitting strategy} 
We used a standard $\chi^2$ minimization,
\begin{equation}
\chi^2 \propto \sum_i \frac{(O_i - M_i)^2}{e_i^2}
,\end{equation}
where $O_i$ is the i-th observations, $e_i$ its associated error, and $M_i$ the prediction from the model. The strategy to compute the overall $\chi^2$, for all observations, necessitates some care. A normal $\chi^2$ would weight each measurement by its error bar. However, when we mix various observables, those that
are present in large numbers are favored compared to scarce ones. A more general approach is to compute a $\chi^2$ by computing the final $\chi^2$ as the average of $\chi^2$ computed for each observable:
\begin{equation}
\chi^2 \propto \sum_j \frac{1}{\mathrm{sizeof(G_j)}} \sum_{i\in G_j} \frac{(O_i - M_i)^2}{e_i^2}
.\end{equation}
This is to ensure that each group $G_j$ of observables contributes equally to the final likelihood estimation: for example, there are usually many more photometric observations than radial velocity or interferometric diameters. We used a Levenberg-Marquardt (LM) least-squares fit based on \texttt{SciPy}\footnote{\url{http://scipy.org/}} \texttt{scipy.optimize.leastsq}. Using the total $\chi^2$ would have given more importance to data in highest numbers. Contrary to the approach taken by \citetads{2012ApJ...748..107P}, we did not fit the zero points of photometric systems, so we do not suffer degeneracy. After we found the best fit, we estimated the uncertainties in the derived parameters by using the covariance matrix around the best-fit solution. 

Another aspect of the fitting process is the phasing of the data. It is known that Cepheids are not perfectly stable pulsators. For example, the slow (compared to the pulsation time) evolution of the star's interior leads to a first-order period change. The amount of linear change is an indicator of the evolutionary stage of the Cepheids and can be computed theoretically (see, for example, \citeads{2014AstL...40..301F}). We allowed the period to change linearly in our model.

\section{Prototypical stars}

Note that the observational data, and best fit model are available in electronic form, as FITS tables.

\subsection{$\delta$ Cep}

$\delta$ Cep is the prototypical Cepheid and has been observed extensively, in particular by optical interferometer. We took the photometry from \citetads{1984ApJS...55..389M}, \citetads{1997PASP..109..645B}, \citetads{1998JAD.....4....3K}, \citetads{2008yCat.2285....0B} and \citetads{2014ApJ...794...80E}. We also added photometric observations from Tycho and Hipparcos from \citetads{1997A&A...323L..61V} and \citetads{1997ESASP1200.....E}. We took the cross-correlation radial velocities from \citetads{1994A&AS..108...25B} and \citeads{2004A&A...415..521S}. The angular diameters are the ones published in \citetads{2005AandA...438L...9M} and \citetads{2006AandA...453..155M}. In addition, to properly interpolate the photospheric models, we adopted a metallically of [Fe/H]=0.06, based on \citeads{2002A&A...381...32A}. We note that the metallicity has a very weak effect on surface brightness values and is undetectable with our data set.

\begin{figure*}
  \centering
  \includegraphics[width=\textwidth]{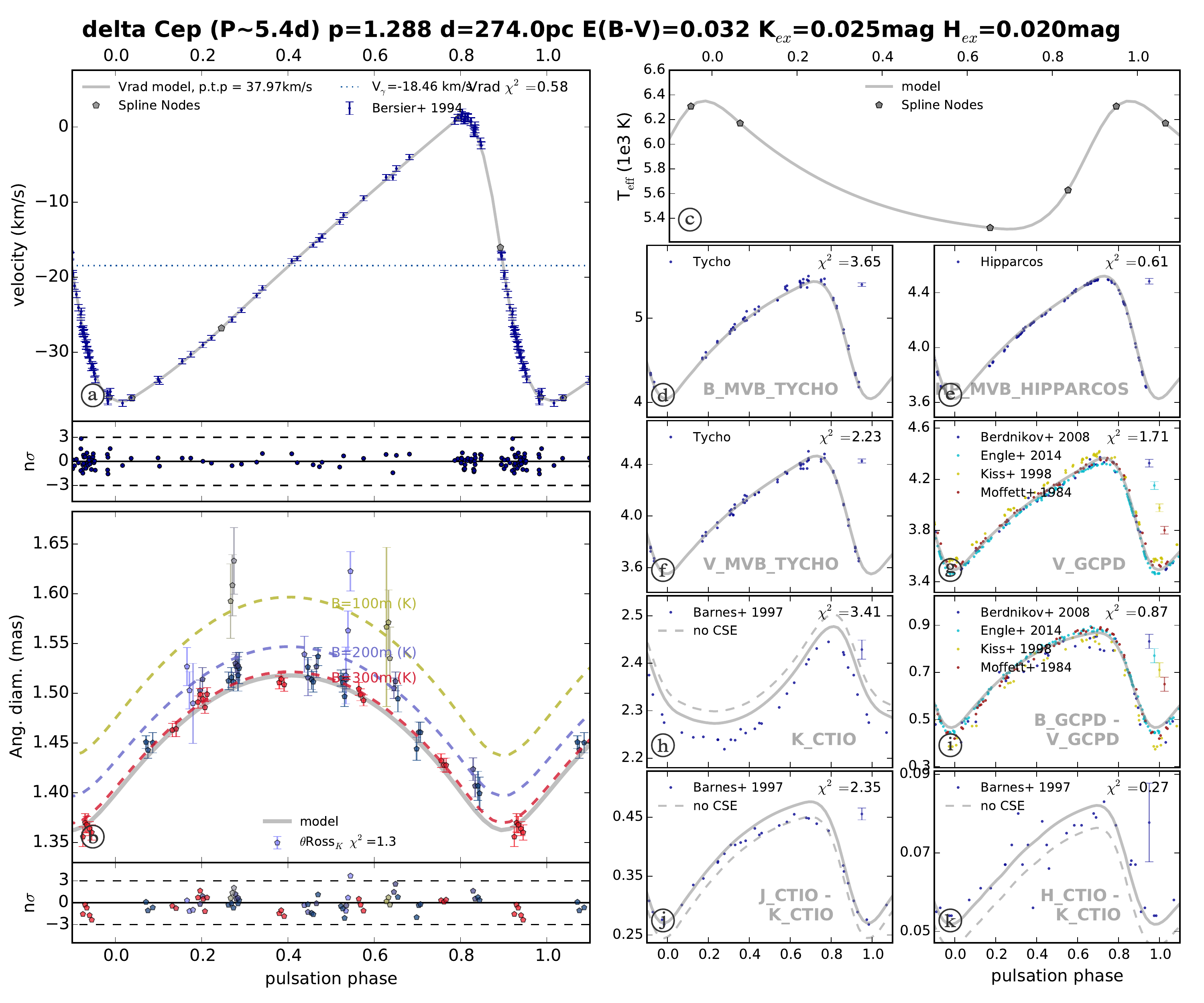}
  \caption{$\delta$~Cep data fit. Various panels show pulsation and radial velocities with spline model and residuals (panel a); angular diameters and residuals, with the baseline color-coded for the data and CSE-biased model ---as a dash line, based on the model shown in Fig.~\ref{Fig:bias}--- (panel b); effective temperatures (panel c); photometric measurements and models (panels d to m) for different photometric bands or colors. Typical error bars are shown on the right side of the plot, below the reduced $\chi^2$ values.}
  \label{Fig:delCep}
\end{figure*}

For the $\chi^2$ averaging, we used four groups of observables: radial velocities (91 measurements) angular diameters (67 measurements), photometric magnitudes (483 measurements), and colors (421 measurements). Error bars for each of these groups were multiplied by $\sim$0.59, $\sim$0.50, $\sim$1.26, and $\sim$1.35, respectively. We show the fit in Fig.~\ref{Fig:delCep}, and the most important parameters are listed in Table~\ref{Tab:delCepFit}.

\begin{table}
  \caption{Parameters of the $\delta$~Cep fit. The quantities with uncertainties are adjusted in the model and the other ones are fixed. We note that the uncertainties are purely statistical and do not take into account systematics, such as the uncertainties on the distance, for example (274$\pm$11~pc, \citeads{2002AJ....124.1695B}).}
  \label{Tab:delCepFit}
  \begin{tabular}{lc}
    \hline
    parameter     &  best fit  \\
\hline
    $\theta_\mathrm{0}^\mathrm{(a)}$ (mas) & $1.420\pm0.009$ \\
    $E(B-V)$ & $0.032\pm0.005_\mathrm{stat.}\pm0.015_\mathrm{sys.}$ \\
    K excess (mag) & $0.025\pm0.002$\\
    H excess (mag) & $0.018\pm0.004$\\
    p-factor &  $1.29\pm0.02$\\
    $\mathrm{MJD}_0^\mathrm{(b)}$ & $48304.7362421$ \\
    period (days)& $5.3662906\pm0.0000061$ \\
    period change (s/yr) & $-0.069\pm 0.033$ \\
    metallicity [Fe/H] & $0.06$ \\
    distance (pc)  & $274$ [fixed]\\
    \hline
    $\chi^2_r$ & 1.7 \\
    \hline
        adopted mass (M$_\odot$) & 4.8 \\
    average radius (R$_\odot$) & 43.0 \\
  \end{tabular}
  \tablefoot{$^\mathrm{(a)}\theta_\mathrm{0}$ is the Rosseland angular diameter
    at phase 0, not the average angular diameter over the pulsation
    cycle; $^\mathrm{(b)}$adjusted so that the bolometric magnitude
    reached minimum at phase 0.}
\end{table}

It is interesting to compare the result we obtain here with that of our previous study, which did not include photometry (\citeads{2005AandA...438L...9M}). The value of the p-factor is very similar: Using only the radial
velocities and angular diameters reported by \citetads{1994A&AS..108...25B}, we found p=1.27$\pm$0.01. The uncertainty was smaller since we took into account correlations in interferometric error bars (using the formalism of \citeads{2003A&A...400.1173P}), which we do not yet have implemented in our current SPIPS fitting algorithm. The actual p-factor uncertainty should take into account the distance uncertainty (0.050), however, which is much larger that the statistical uncertainty (0.020).

The CSE is noticeable in the interferometric data as a bias affecting the angular diameter measured at the shortest baselines. \citetads{2006AandA...453..155M} did not fit the excess, but rather compared the fit using a simple star model to a fit using the model we fitted on another Cepheid (Polaris), for which we had extended baseline coverage. At the time, we used a 1.5\% excess (0.016 mag). In the case of SPIPS, we have photometric data that  allow anchoring the model and allow using the CSE contribution as a free parameter. Thanks to this, we confirmed the infrared excess and estimated it to be 0.025$\pm$0.002 mag in K band. We also let the H excess free
to vary to fit the photometry and found it to be 0.018$\pm$0.004. This latter is solely based on the photometric measurements.

The good agreement with all the observables is remarkable and increases our confidence in the method. In particular, our SPIPS modeling is able to combine all data and does not show apparent discrepancies between optical interferometry and IRSB, such as noted by \citetads{2012AandA...543A..55N}. Admittedly, we added the complexity of having an infrared excess, which probably explains the discrepancy (which \citeads{2012AandA...543A..55N} did not take into account). One could argue that the K-band magnitudes do not agree the best agree in our fit (Fig.~\ref{Fig:delCep}, panel 'h'). We also performed a fit using only photometric measurements (omitting our interferometric measurements) and found the p-factor to be $1.29\pm0.06,$ which, apart from the poorer statistical uncertainty, agrees perfectly well with our fit using optical interferometry. The K excess was also let free in the photometric fit, and its value was found to be $0.010\pm0.004$ magnitude.

Additionally, the period change ($-0.07\pm0.03$~s/yr) is found to agree well with the recent estimate by \citeads{2014ApJ...794...80E}, even though these authors have a much greater accuracy ($-0.1006\pm0.0002$~s/yr).

\subsection{$\eta$ Aql}

$\eta$~Aql is another important prototypical Cepheid because of its proximity (and hence large apparent size), which makes it accessible to optical interferometry. We observed  $\eta$~Aql in July 2006, using the FLUOR instrument (\citeads{2003SPIE.4838..280C}) at the CHARA Array. We used the same data reduction approach as in previous works, in particular the $\delta$~Cep data used in the previous section. 

We took photometry from \citetads{1984ApJS...54..547W}, \citetads{1984ApJS...55..389M}, \citetads{1997PASP..109..645B}, \citetads{1998JAD.....4....3K}, \citetads{2008yCat.2285....0B}. Photometric measurements in the Walraven system were taken from \citetads{1976A&AS...24..413P}. We also added photometric observations from Tycho and Hipparcos from \citetads{1997A&A...323L..61V} and \citetads{1997ESASP1200.....E}. Radial velocities were taken from \citetads{2005ApJS..156..227B} and \citetads{1998JAD.....4....3K}. Finally, we also took additional angular diameter measurements: H band long-baseline measurements from \citetads{2002ApJ...573..330L} and short-baseline K-band measurements from \citetads{2004A&A...416..941K}. We adopted a metallically of [Fe/H]=0.05 , based on \citeads{2002A&A...381...32A}.

\begin{figure*}
  \centering
  \includegraphics[width=\textwidth]{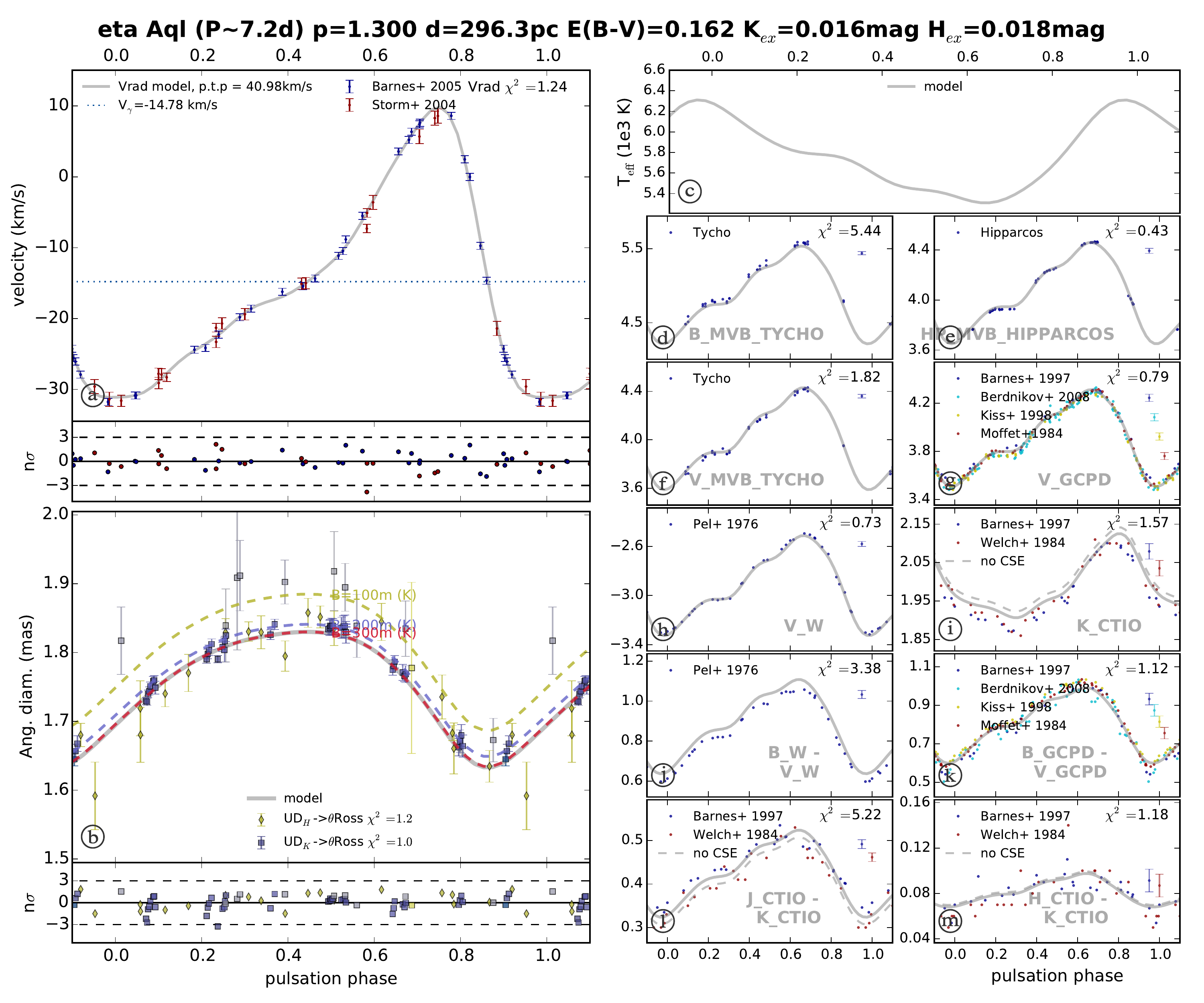}
  \caption{$\eta$~Aql fit. Various panels show pulsation and radial velocities with spline model and residuals (panel a); angular diameters and residuals, with the baseline color-coded for the data and CSE-biased model --- as a dash line, based on the model shown in Fig.~\ref{Fig:bias}--- (panel b); effective temperatures (panel c); photometric measurements and model (panels d to o) for different photometric bands or colors. Typical error bars are shown on the right side of the plot, below the reduced $\chi^2$ values.}
  \label{Fig:etaAql}%
\end{figure*}

\begin{table}
  \caption{Parameters of the $\eta$~Aql fit. The quantities with uncertainties are adjusted in the model and the other ones are fixed.}
  \label{Tab:etaAqlFit}
  \begin{tabular}{lc}
    \hline
    parameter     &  best fit \\
    \hline
    $\theta_\mathrm{0}^\mathrm{(a)}$ (mas)& $1.694\pm0.002$ \\
    $E(B-V)$ & $0.161\pm0.005_\mathrm{stat.}\pm0.015_\mathrm{sys.}$ \\
    K excess (mag) & $0.018\pm0.002$\\
    H excess (mag) & $0.016\pm0.003$\\
        p-factor & 1.30 [fixed]\\
    distance (pc)  & $396\pm6$\\
    $\mathrm{MJD}_0^\mathrm{(b)}$ & $48069.3905$ \\
    period (days)&  $7.176841\pm0.000012$ \\
    period change (s/yr) & $0.18\pm0.07$ \\
    metallicity [Fe/H]& $0.05$\\
    \hline
    reduced $\chi^2$ & 2.3\\
    \hline
        adopted mass (M$_\odot$) & 6.3 \\
    average radius (R$_\odot$) & 57.6 \\
  \end{tabular}
  \tablefoot{$^\mathrm{(a)}\theta_\mathrm{0}$ is the Rosseland angular diameter
    at phase 0, not the average angular diameter over the pulsation
    cycle $<\theta>$; $^\mathrm{(b)}$adjusted so that the bolometric magnitude
    reached minimum at phase 0.}
\end{table}

The results of the fit are presented in Fig.~\ref{Fig:etaAql} and Table~\ref{Tab:etaAqlFit}. As for $\delta$~Cep, we applied a correction factor to the error bars to equally weight the four following groups: radial velocities (57 measurements, 0.5 factor), angular diameter (70 measurements, 0.55 factor), photometric magnitudes (377 measurements, 1.3 factor), and photometric colors (432 measurements, 1.35 factor).

We detect a slight H- and K-band infrared excess ($0.016\pm0.003$ and $0.018\pm0.002,$ respectively). Like $\delta$~Cep, this is allowed by the combination of infrared photometry and infrared interferometric angular diameters. 

Regarding the accuracy of E(B-V), \citeads{2007MNRAS.377..147L} reported 0.126 and also quoted an older value of 0.143 (\citeads{1985MNRAS.212..879C}), as well as 0.138 (metallicity corrected, computed by the software 'BELRED'). \citetads{2008A&A...488...25G} quoted $0.130\pm0.009$. \citetads{2011AandA...534A..94S} used 0.129. Our estimate is in this range, at $0.161\pm0.005$, on the redder side. The statistical uncertainty we obtain, $\pm0.005$, is underestimated because we did not take into account the fact that all photometric measurements in a same band and from a same source share a common error, namely the zero point and the photometric calibrators. If we perform a Jack-knife resampling, removing one set of photometric measurements every time, the uncertainty on E(B-V) increases by a factor of 3, to $\pm0.015$.

Regarding the distance, $\eta$~Aql appears in Table 5 of \citetads{2008A&A...488...25G} with a distance of $261\pm6\pm7$pc for p=1.321 ($d/p=198\pm5\pm4$pc). \citetads{2011AandA...534A..94S} determined a distance of $255\pm5$pc using IRSB method, for p=1.39 ($d/p=183\pm4$pc). Using a subset of data we used, \citetads{2002ApJ...573..330L} obtained d=320$\pm$32pc with p=1.43 (d/p=223$\pm$22pc). Our method gives a distance of $296\pm5$pc ($d/p=228\pm4$pc), which is not consistent with \citetads{2011AandA...534A..94S}. We note that our uncertainty is on the same order as that of \citetads{2011AandA...534A..94S}, and surprisingly, they used only radial velocity and two-band photometry. If we restrict ourselves to IRSB data (radial velocities and V, K photometry), our fit leads to $\pm15$pc. Since we cannot fit E(B-V) (because of the degeneracy with T$_\mathrm{eff}$), we should estimate the sensitivity of the distance estimate to change in E(B-V). We computed that decreasing E(B-V) by 0.05 leads to a distance 4pc smaller. In other words, restricting our data set to the IRSB method leads to similar distances. The reason why we find an uncertainty in the estimated distance three times larger than \citetads{2011AandA...534A..94S} is the following: we suspect that since we fitted all parameters at once (radial velocity profile, T$_\mathrm{eff}$ profile, distance, etc.), our uncertainties are more realistic. If we keep our $\eta$~Aql model and only use the IRSB dataset, and if we assume that we know everything in the model except for the distance and only adjust for this parameter, the uncertainty decreases to $\pm5$pc, which is the claim of \citetads{2011AandA...534A..94S}. In other words, our analysis of $\eta$~Aql is a perfect example of why fitting all parameters at the same time provides more realistic uncertainties.

\section{Conclusions}
Our model makes many simplistic assumptions about Cepheids, most of which are known to be incorrect at a certain level. However, in the context of the parallax-of-pulsation distance estimation, our approach is more complete than most (if not all) implementations that are variations of the Baade-Wesselink method (BWM): 1) we include all possible observables, including redundant ones, and 2) we use observation modeling based on a physical model (as opposed to ad-hoc parameters, such as the surface brightness relations). Our implementation includes the traditional BWM, if one restricts the input data set. Using our modeling, we address some shortcomings of the BWM:
\begin{itemize}
\item We adopted an approach of modeling the observables rather than using ill-defined corrective factors. For example, we used modeled interferometric visibility profiles to compute the interferometric bias $\theta_\mathrm{UD}/\theta_\mathrm{Ross.}$ whereas it is traditionally derived for brightness profile fits to analytical functions. We still make use of the projection factor, but we are working on a spectral synthesis modeling to allow us to use a consistent pulsation velocity estimation.
\item We used atmospheric models (ATLAS9 in our case) to compute synthetic photometry. This works very well, as proven by the
agreement with interferometric angular diameters on our two prototypical stars. We note that the resulting surface brightness relation cannot be approximated by a linear function of the effective temperature (or color), as is done with a traditional implementation of the BWM. Because the BWM lacks redundancy in the dataset it uses, this shortcoming cannot be detected and propagates as a color bias on the distances.
\item Circumstellar envelopes (CSE) are consistently taken into account in the near-infrared photometry and optical interferometric diameters.
\item Reddening is fitted from the data in a self-consistent way. Conversely, BWM uses an E(B-V) that was determined for a certain reddening law and often applies it using another reddening law. Our method does not suffer from this bias.\item Our approach permits very good phasing of data, even taken at different epochs. Not only does it improve the accuracy of the distance determination (because poorly phased data often have underestimated amplitude), it also allows us to study the period change of Cepheids.
\item Fitting all parameters at once realistically estimates the statistical uncertainties, as opposed to a method that would fit consecutive sets of parameters. For example, if the analytical radial velocity function is fitted
first in an implementation of the BMW and then the analytical variations of angular diameters, followed by the distance alone as the ratio between the two, the uncertainty of the distance would not account for the other uncertainties and would likely be underestimated by a factor as large as 3.
\end{itemize}
All this should come as a warning to studies using only two bands: their distance (or p-factor) determinations probably have systematic errors that are hard to estimate without using a method like the one we have presented. Even then, their statistical uncertainties might very well be underestimated by a large factor.
We applied the method to $\delta$~Cep and $\eta$~Aql. For $\delta$~Cep we confirm our formerly published values for the p-factor of 1.28$\pm$0.06, accounting for the uncertainty of the distance by \citeads{2002AJ....124.1695B} of $274\pm11$~pc. For $\eta$~Aql, we estimated its biased distance to be $d/p=228\pm4$pc, leading to $d=296\pm5$pc assuming p=1.30. In both cases, our models reproduced all the available data (about a thousand observations in each case), in a self-consistent way. In the near future, we will continue our work by systematically studying Cepheids for which large datasets are available.

\begin{acknowledgements}
We would like to thank the referee, Hilding Neilson, for his work that led to a much improved manuscript, as well as for providing additional insights to the use of SATLAS models described in the present work.

This research has made use of the Spanish Virtual Observatory supported from the Spanish MEC through grant AyA2008-02156. This research has made use of the VizieR catalog access tool and SIMDAD database, operated at CDS, Strasbourg, France. 

A.G. acknowledges support from FONDECYT grant 3130361. P.K and J.B acknowledge financial support from the "Programme National de Physique Stellaire" (PNPS) of CNRS/INSU, France, and the ECOS/Conicyt grant C13U01. 

The CHARA Array is funded by the National Science Foundation through NSF grants AST-0908253 and AST-1211129, and by the Georgia State University through the College of Arts and Sciences. 

STR acknowledges support by NASA through grant number HST-GO-12610.001-A from the Space Telescope Science Institute, which is operated by AURA, Inc., under NASA contract NAS 5-26555.
\end{acknowledgements}

\bibliographystyle{aa} 
\bibliography{biblio} 

\end{document}